\documentclass[prc,aps,eqsecnum,amssymb,floatfix]{revtex4}
\def\tri{{{}^3{\rm H}}}

\def\be{\begin{equation}}
\def\ee{\end{equation}}
\def\bea{\begin{eqnarray*}}
\def\eea{\end{eqnarray*}}

\def\bi{\begin{itemize}}
\def\ei{\end{itemize}}

\def\m{\phantom{-}}

\usepackage{amsfonts}
\usepackage{amssymb}
\usepackage{graphicx}
\begin{document}

\title{Neutron-triton elastic scattering}

\author{M. Viviani$^1$, A. Kievsky$^2$, L. Girlanda$^{2,1}$,
        and L. E. Marcucci$^{2,1}$}
\affiliation{(1) INFN, Sezione di Pisa, Largo Pontecorvo, 3, 56127 Pisa
  (Italy) \\
(2) Phys. Dept., University of Pisa, Largo Pontecorvo, 3, 56127 Pisa (Italy)}



\date{\today}

\begin{abstract}
The Kohn variational principle and the
hyperspherical harmonics technique are applied to study the
$n-\tri$ elastic scattering at low energies.
In this contribution the first results obtained using a non-local realistic
interaction derived from the chiral perturbation theory are reported. They
are found to be in good agreement with those obtained solving the
Faddeev-Yakubovsky equations. The calculated total and differential cross
sections are compared with the available experimental data.  The effect of
including a three-nucleon interaction is also discussed.
\end{abstract}

\maketitle

\section{Introduction}
\label{sec:intro}

In the last few years the scattering of nucleons by deuterons has been
the subject of a large  number of investigations. This scattering problem is in 
fact a very useful tool for testing the accuracy of our present knowledge of
the nucleon--nucleon (NN) and three nucleon (3N) interactions. Noticeable
progress has been achieved, but a number of relevant disagreements 
between theoretical predictions and experimental results still remains to be
solved~\cite{GWHKG96,KRTV96}.   

It is therefore of interest to extend the above mentioned analysis 
to four nucleon scattering processes. In this case, an important goal for both 
theoretical and experimental analysis is to reach a precision comparable to
that  achieved in the $N-d$ case. This is particularly challenging from the
theoretical point of view, since the study of $A=4$ systems is noticeably more
complicated than the $ A=3$ one. Recently, accurate calculations of four-body
scattering observables have been achieved in the framework of the
Faddeev-Yakubovsky (FY) equations~\cite{DF07}, solved in momentum
space, and treating the long-range Coulomb interaction using the
screening-renormalization method~\cite{DF07b,DF07c}.

In this contribution, the four-body scattering problem is solved using the 
Kohn variational method and expanding the internal part of the wave function
in terms of the hyperspherical harmonic (HH) functions. Previous applications of
this method~\cite{VKR98,Vea01,Lea05} were limited so far to consider only
local potentials, as the Argonne V18~\cite{AV18} NN potential. Recently, 
for bound-states, the HH method has been extended to treat also non-local
potentials, given either in coordinate- or momentum-space~\cite{Vea06}. Here,
we report the first application of the HH method to the four-body scattering
problem with non-local potentials.

The potential used in this paper is the N3LO-Idaho model by Entem \&
Machleidt~\cite{EM03}, with cutoff $\Lambda=500$ MeV.  
This potential has been derived using an  effective field theory approach
and the chiral perturbation theory up to next-to-next-to-next-to-leading order.
We have also performed calculations by adding to the N3LO-Idaho potential a 
3N interaction, derived at next-to-next-to leading order (N2LO)
in Ref.~\cite{N07} (N3LO-Idaho/N2LO interaction model). The two free parameters in this
N2LO 3N potential have been chosen from the combination that reproduces
the $A=3,4$ binding energies~\cite{N07}. The development of a 3N interaction
including  N3LO contribution is still under progress~\cite{Bea07}.

\begin{table}[b]
\begin{tabular}{l@{$\qquad$} c@{$\quad$}c@{$\quad$}|
                l@{$\quad$}c@{$\quad$}c}
\hline
Phase-shift & HH  & FY   & Phase-shift & HH  & FY   \\
\hline
${}^1S_0$  & $-69.3$  & $-69.1$  & ${}^3P_0$  & $23.2$  & $23.3$   \\
\hline
${}^3S_1$  & $-61.4$  & $-61.2$   & ${}^1P_1$  & $22.7$  & $22.5$   \\
${}^3D_1$  & $-1.14$  & $-1.10$   & ${}^3P_1$  & $44.4$  & $44.5$   \\
$\epsilon$ & $\m0.77$ & $\m0.80$  & $\epsilon$  & $9.80$  & $9.64$   \\
\hline
${}^1D_2$  & $-1.72$  & $-1.90$   & ${}^3P_2$  & $48.4$  & $48.7$   \\
${}^3D_2$  & $-0.94$  & $-1.01$   & ${}^3F_2$  & $0.07$  & $0.09$   \\
$\epsilon$  & $\m2.74$  & $\m2.81$  & $\epsilon$  & $1.24$  & $1.26$   \\
\hline
\end{tabular}
\caption[Table]{\label{table:comp}
Phase-shift and mixing angle parameters for $n-\tri$ elastic
scattering at incident neutron energy $E_n=4$ MeV calculated using the
N3LO-Idaho potential. 
The values reported in the columns labeled HH have been
obtained using the HH expansion and the Kohn variational principle, whereas
those reported in the columns labeled FY by solving the FY
equations~\protect\cite{DF07}. 
}
\end{table}

This paper is organized as follows. In Section~\ref{sec:comp}, a comparison
between HH and FY calculations is reported. We have performed this comparison
for the N3LO-Idaho potential for incident neutron energy $E_n=4$ MeV. Finally, in
Section~\ref{sec:res}, the theoretical calculations are compared with the
available experimental data.

\section{Comparison between HH and FY results}
\label{sec:comp}

The calculated phase-shift and mixing angle parameters for $n-\tri$ elastic
scattering at $E_n=4$ MeV using the  N3LO-Idaho potential are reported in
Table~\ref{table:comp}. The values reported in the columns labeled HH have been
obtained using the HH expansion and the Kohn variational principle, whereas
those reported in the columns labeled FY by solving the FY
equations~\cite{DF07b}. As can be seen, there is a good overall agreement
between the results of the two calculations.

\section{Results}
\label{sec:res}

\begin{figure}
 \includegraphics[height=8cm]{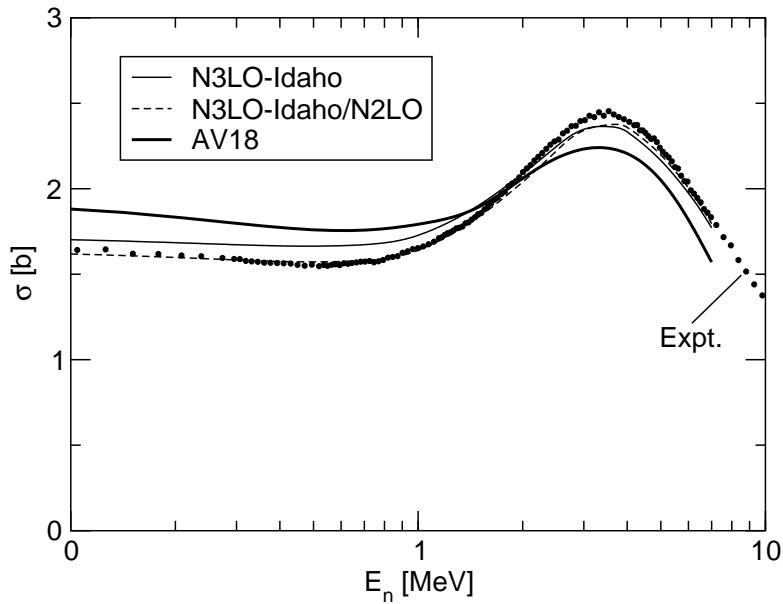}
\caption{$n-\tri$ total cross sections calculated with the AV18 (thick solid
  line), N3LO-Idaho (solid line), and the N3LO-Idaho/N2LO (dashed line) as
  function of the incident neutron energy $E_n$. The experimental
  data are form Ref.~\protect\cite{PBS80}.}
\label{fig:tcs}
\end{figure}

The preliminary results for the $n-\tri$ total cross section calculated with
the considered  potential models are reported in Figure~\ref{fig:tcs}. As already
known, the calculated cross section with the AV18 potential overpredicts the
experimental data at low energies, and is well under the data in
the peak region~\cite{Lea05,DF07b}. The problem at low energies is cured when
the Urbana-IX 3N force~\cite{UIX} is considered~\cite{Lea05}. In the peak region
the inclusion of this 3N force slightly decreases the cross section,
increasing the disagreement with the data. On the other hands, using the N3LO-Idaho
a better agreement with experimental data is found~\cite{DF07b}.
Including the N2LO 3N force, there is now a perfect
agreement at low energy (in particular, in the minimum around $E_n=1$ MeV). 
Also in the peak region a slight better agreement is observed. The origin of
the remaining discrepancy is unclear, but it could be related to parts of
3N interaction not yet considered.

\begin{figure}
 \includegraphics[height=7cm]{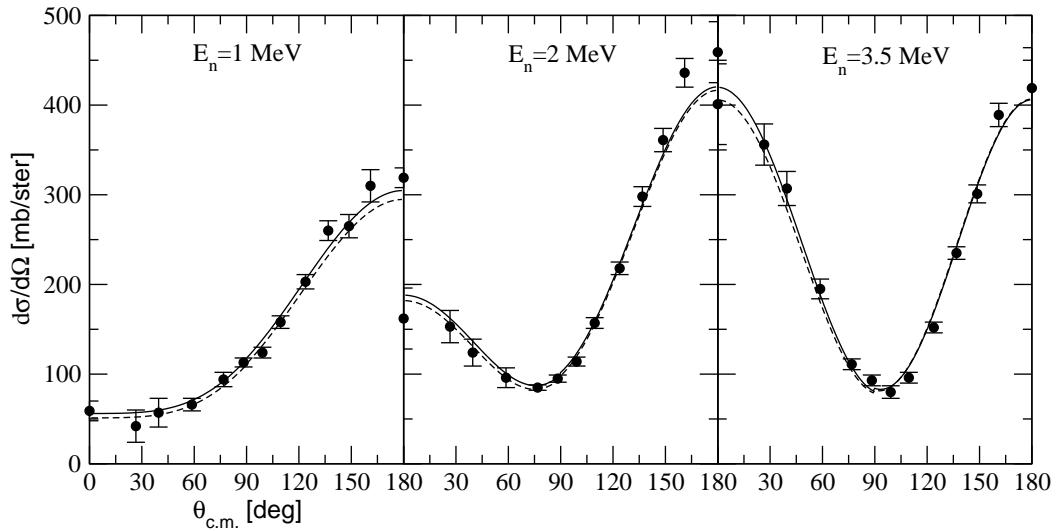}
 \caption{$n-\tri$ differential cross sections calculated with the
  N3LO-Idaho (solid line) and the N3LO-Idaho/N2LO (dashed line) interaction
  models for three different incident neutron energies. The experimental data
  are from Ref.~\protect\cite{SCS60}. } 
\label{fig:dcs}
\end{figure}

The quality of the agreement can be also seen by comparing the theoretical
and experimental differential cross sections, avaliable at
$E_n=1$, $2$, and $3.5$ MeV.

\end{document}